\documentclass{sig-alternate}
\newcommand{\mo}{{\it mathoverflow }}
\begin{document}
%
% --- Author Metadata here ---
\conferenceinfo{SOHUMAN 2013}{Paris, France}
%\CopyrightYear{2007} % Allows default copyright year (20XX) to be over-ridden - IF NEED BE.
%\crdata{0-12345-67-8/90/01}  % Allows default copyright data (0-89791-88-6/97/05) to be over-ridden - IF NEED BE.
% --- End of Author Metadata ---
\title{What does mathoverflow tell us about the production of mathematics?
%\title{Analysing mathematical question answering  to lay the groundwork for a mathematics social machine
%\titlenote{(Produces the permission block, and
%copyright information). For use with
%SIG-ALTERNATE.CLS. Supported by ACM.}}
%\subtitle{[Extended Abstract]
%\titlenote{A full version of this paper is available as
%\textit{Author's Guide to Preparing ACM SIG Proceedings Using
%\LaTeX$2_\epsilon$\ and BibTeX} at
%\texttt{www.acm.org/eaddress.htm}}
}

\numberofauthors{2} %  in this sample file, there are a *total*
\author{
% 1st. author
\alignauthor
Ursula Martin\\
       \affaddr{Queen Mary University of London}\\
       \email{Ursula.Martin@eecs.qmul.ac.uk}
% 2nd. author
\alignauthor
Alison Pease\\
      \affaddr{Queen Mary University of London}\\
       \email{Alison.Pease@eecs.qmul.ac.uk}
}
% There's nothing stopping you putting the seventh, eighth, etc.
% author on the opening page (as the 'third row') but we ask,
% for aesthetic reasons that you place these 'additional authors'
% in the \additional authors block, viz.
\date{1st February 2013}
% Just remember to make sure that the TOTAL number of authors
% is the number that will appear on the first page PLUS the
% number that will appear in the \additionalauthors section.

%refs 56	
%examples 34	
%error 37
%
%rules say "question with clear answer"

\maketitle
\begin{abstract}
The highest level of mathematics research is traditionally seen as a solitary activity. Yet new innovations by mathematicians themselves are starting to harness the power of social computation to create new modes of mathematical production. We study the effectiveness of one such system, and make proposals for enhancement, drawing on AI and  computer based mathematics. We analyse the content of a sample of questions and responses in the community question answering system for research mathematicians, {\it mathoverflow}. We find that {\it mathoverflow} is very effective, with 90\% of our sample of questions answered completely or in part. A typical response is an informal dialogue, allowing error and speculation, rather than rigorous mathematical argument: 37\% of our sample discussions acknowledged error. Responses typically present information known to the respondent, and readily checked by other users: thus the effectiveness of {\it mathoverflow} comes from information sharing. We conclude that extending and  the power and reach of {\it mathoverflow} through a combination of people and machines raises new challenges  for artificial intelligence and computational mathematics, in particular how to handle error, analogy and informal reasoning.
%
%With the long term goal of building social machines for mathematics, we analyse the content of a sample of questions and responses in the community question answering system, {\it mathoverflow}. We find that \mo is very effective,  with 90\% of our sample of questions answered completely or in part.  A typical response is an informal dialogue, allowing error and speculation, rather than rigorous mathematical argument: 37\% of our sample discussions acknowledged error. Responses typically present information known to the respondent, and readily checked by other users: thus the effectiveness of \mo comes from information sharing. We conclude that building social machines which extend the power and reach of \mo raises challenges that go well beyond traditional computer mathematics.
\end{abstract}



\section{Introduction}
%Social machines, introduced by Berners-Lee and Fischetti in \cite{Berners-Lee:2001fk}, and amplified by Hendler and Berners-Lee \cite{hendler2010semantic}, combine people and computers into a single problem-solving entity ``in which the people do the creative work and the machine does the administration.'' 

Our long term goal is to  advance mathematics research through combining  social computation with more established computer support: computational mathematics, artificial intelligence and mathematical knowledge management. 

In this paper we study a specific instantiation of collective intelligence and social computing, the  mathematics community question answering system \mo.  We seek to understand whether and how it is an effective problem solving mechanism, and how we might combine it with other computer based approaches to mathematics, to build future more effective systems.

We analyse a sample of questions in  \mo to understand the structure and mathematical content of questions and responses in a technical research domain.  We find that \mo is very effective,  with 90\% of our sample of questions either answered completely or in part.  We developed a typology of questions, and showed that  majority of questions (64\%) asked for factual information: conjectures;  proofs; examples;  formulas or references. The remainder (excluding the 2\% categorised as ``Other'')  were more open ended, seeking to understand  phenomena, clarify difficulties, or understand motivation.  

Mathematics is typically thought of as a formal logical activity: however a typical interaction on \mo is an informal dialogue, rather than rigorous steps of correct mathematical inference. Speculative statements are made, and errors corrected when pointed out, a pattern found in 37\% of our sample.  Responses typically present information known to the respondent, and readily checked by other users. 56\% of answers refer to the existing research literature, and 34\% supply  examples: in both cases the responder may be pointing out consequences or properties of the material that would not be apparent to a less expert reader.  Thus the power of \mo comes from developing collective intelligence through sharing information and understanding. 

Future social computation for mathematics which extend the power and reach of \mo will need computer support that goes well beyond established computer support: computational mathematics, artificial intelligence and mathematical knowledge management, to support cognitive approaches encompassing informal reasoning, and bodies of individual user knowledge. 

\section{Background}
%The research agenda set out in  \cite{hendler2010semantic} is ambitious,  drawing attention to challenges for artificial intelligence in developing the tools that will allow users to develop their own social machines. These include pragmatic means of developing ontologies that turn ``messy human knowledge'' into a usable information space;  providing mechanisms for treating social expectations (e.g. privacy) and legal rules as first class objects; reasoning mechanisms for open systems that must deal with scale and inconsistency; dealing effectively with context; and developing notions of information accountability.

Research mathematics provides a novel domain for social computation research.  
 For centuries, the highest level of mathematics has been seen as an isolated creative activity, to produce  a proof for review and acceptance by research peers.  While computation has transformed other sciences, its use in advanced mathematics, for example in computer support of mathematical proof, has been subject to vigorous debate \cite{martin1999computers}.   
Recent significant advance is starting to change perceptions: for example Hales has almost completed a ten-year computer proof of the Kepler conjecture \cite{hales2005proof}.

The mathematical community were ``early adopters'' of the internet for disseminating papers, sharing data, and blogging, and in recent years have developed their own social computation systems for ``crowdsourcing'' (albeit among a highly specialised crowd) the production of mathematics.  

For example, the widely used ``Online Encyclopaedia of Integer Sequences''  (www.oeis.org), given a few digits of input, proposes sequences which match it, through invoking subtle pattern matching against over 220,000 user-provided sequences: so for example user input of (3 1 4 1) returns $\pi$, and 556 other possibilities, supported by links to the mathematical literature. Viewed within the framework of social computation, it involves users with queries or proposed new entries; a wiki for discussions; volunteers curating  the system; governance and funding mechanisms through a trust; alongside traditional computer support for a database, matching engine and web interface, with links to other data sources, such as research papers. While anyone can use the system, proposing a new sequence requires registration and a short CV, which is public, serving as a reputation system.   

A particularly striking example of a mathematics social machine is  {\it polymath}. Developed by Fields Medal\footnote{The mathematics Nobel Prize} winner Sir Tim Gowers, with clear ground rules to encourage users to show their working, be polite and constructive and so on, this uses a wiki for collaboration among mathematicians from different backgrounds  to develop proofs, and has led  to major advances and publications containing new results \cite{gowers2009massively}. {\it polymath} discussions also provide a rich data source for the analysis of mathematical practice \cite{Pease12}, giving insights into the roles of conjecture, concept formation, examples and error alongside proof steps  in the production of mathematics, and insight into possible future  {\it polymath} social computation.  

Discussion fora for research mathematics have evolved from the early newsnet newsgroups to modern systems based on the {\it stackexchange} architecture, which allow rapid  informal interaction and problem solving. In three years {\it mathoverflow.net} has accumulated 16,000 users and hosted 27,000  conversations. The highly technical nature of research mathematics means that,  in contrast to activities like GalaxyZoo, this is not currently an endeavour accessible to the public at large: a separate site  {\it math.stackexchange.com} is a broader question and answer site ``for people studying math at any level and professionals in related fields''.

WIth \mo, house rules give detailed guidance, and stress clarity, precision, and asking questions with a clear answer. Moderation is fairly tight, and  some complain it constrains discussion.  The design of such systems has been subject to considerable analysis by the designers and users, and {\it meta.mathoverflow} contains many reflective discussions.  A key element of the success of the system user ratings of questions and responses, which combine to form reputation ratings for users. These have been studied by psychologists Tausczik and Pennebaker \cite{tausczik2011predicting, tausczik2012participation}, who  concluded that \mo reputations offline (assessed by numbers of papers published) and in \mo  were consistently and independently related to the \mo ratings of authors' submissions, and that while more experienced contributors were more likely to be motivated by a desire to help others, all users were motivated by building their \mo  reputation. The work highlights the importance of user identity and community roles, opening up further questions - for example whether users felt themselves becoming more competent as a result of engaging, or if the community organizes itself into learners and teachers in different domains.

In this paper, rather than study such user behaviour, we study the mathematical content of \mo questions and responses.  We chose the subdomain of group theory \footnote{The first author is a  group theorist \cite{helleloid2007automorphism}}: at the time of writing (March 2013) around  2000 of the \mo questions are tagged  ``group theory'', putting it in the top 5 topic-specific tags.  

A group is, roughly speaking, the set of symmetries of an object, and the field emerged in the nineteenth  century, through the systematic study of roots of equations triggered by the work of Galois, and continues to provide a surprising and challenging abstract domain which underlies other parts of mathematics, such as number theory and topology, with practical applications in areas such as cryptography and physics. Its greatest intellectual  achievement is the classification of finite simple groups \cite{gorenstein1982finite}, the basic ``building blocks'' of all finite groups, a result taking many thousands of journal pages over 30 years. The field has a  well-developed tradition of computer support and online resources, making it a good candidate for further social computation  support: early  stand alone programmes  have developed into  widely used software such as GAP \cite{GAP4}, incorporating many specialist algorithms, and exhaustive  online data sources.

\section{What do the questions tell us?}
A  \mo user asking a question seeks assistance from other users.  In their study of intent in users of Yahoo! Answers and MSN QnA,  Rodrigues and Milic-Frayling \cite{mendes2009socializing} constructed a  typology of eight kinds of question: { Factual Information; General Advice; Personal Advice; General Opinion; Personal Opinion;  Chatting; Entertainment; Other} 
%
%which they characterise along three broad axes of 
% {\bf Personal versus General perspective};
% {\bf Community versus Individual issue}; {\bf  Social versus Non-social intent}.
%This 
which refined the earlier classification used by \cite{adamic2008knowledge}, whose typology distinguishes between questions seeking factual information, advice, or opinions. 
%The question formats of \mo are restricted by the house rules described above.  A preliminary inspection of our question data suggested that the majority of the questions fell into the  categories characterised by  \cite{mendes2009socializing} as:
%\begin{itemize}
%\item{\bf Factual Information} --- the question is a request for factual information or a source of information. It may require more or less expertise to answer.
%\item{\bf General Advice} --- request for advice or recommendation about general or personal issues, but provided from an objective stance, involving verifiable facts. The user intent Êis to solve a problem, make a decision or carry out an action. 
%\end{itemize}
%The analysis of \cite{mendes2009socializing} focusses on two aspects of question asking, the {\it intention of the users} and the {\it type of information requested}.  While the house rules of \mo encourage users to give some background and explain why they are asking for information, our preliminary examination suggested that this  is not often provided. Hence we sought an alternative typology that more precisely captured the kinds of information and advice that were sought by specialist mathematicians asking questions of each other. 

The social sciences offer a variety of techniques for qualitative data analysis. Following  \cite{mendes2009socializing}, we use Thomas's  {\it Undirected Inductive Coding} (UIC) method \cite{thomas2006general}, which, rather than working top-down with a fixed set of codes, allows the user to generate codes bottom-up, starting with a small set of codes and generating a new code whenever a data item cannot be covered by existing codes. The process of code reduction then combines the codes into broader categories.

We analysed a sample of 100 questions (drawn from April 2011 and July 2010 to obtain a spread) using UIC, and refined the results of \cite{mendes2009socializing} to develop a typology based on  the kinds of questions being asked.  Excluding the 2\% categorised as ``Other'', our specialised domain fell into only two of the eight categories of \cite{mendes2009socializing}: viz 64\% were Factual Information (conjectures;  proofs; examples;  formulas or references) and 34\% were General Advice (more open ended questions, asking for help in understanding  phenomena, clarifying difficulties, or  motivation).

In general, these corresponded to terms like ``example'', ``formula'' and so on, but the question format could be quite varied. An alternative method of analysis might consider looking for specific ``question words'' such as ``Why'', an approach adopted in the emerging field of experimental philosophy \cite{Overton12}. However this leads to many philosophical subtleties, and was less suited to our more pragmatic primary goal of identifying possible routes for computer support for {\it mathoverflow}.

 Our typology refined that of  \cite{mendes2009socializing}  to encompass: 
\begin{description}
\item [Conjecture 36\%] --- asks if a mathematical statement  is true. May ask directly ``Is it true that'' or ask under what circumstances a statement  is true.
\item [What is this 28\%] --- describes a mathematical object or phenomenon and asks what is known about it.
\item [Example 14\%] ---  asks for examples of a phenomenon or an object with particular properties \item [ 
Formula 5\%] ---   ask for an explicit formula or computation technique.
\item [Different proof 5\%] --- asks if there is an alternative to a known proof. In particular, since our sample concerns the field of group theory, a number of questions concern whether a certain result can be proved without recourse to the classification of finite simple groups.
\item [
Reference 4\%] --- asks for a reference for something the questioner believes to be already in the literature 
\item [
Perplexed 3\%] --- ask for help in understanding a phenomenon or difficulty. A typical  question in this area might concern why accounts from two different sources (for example Wikipedia and a published paper)  seem to contradict each other.
\item [ 
Motivation 3\%] ---  asks for motivation or background. A typical question might ask why something is true or interesting, or has been approached historically in a particular way.
\item [ Other 2\%] --- closed by moderators as out of scope,  duplicates etc.
\end{description}
\section{What do the responses tell us?}

Rather than repeat a fine-grained UIC methodology, we looked in  this initial analysis for broad phenomena in the structure of the successful responses. 

  \mo is very effective,  with 90\% of our sample successful, in that they received responses that the questioner flagged as an ``answer'', of which 78\% were reasonable answers to the original question, and a further 12\% were partial or helpful responses that moved knowledge forward in some way.   The high success rate suggests  that, of the infinity of possible mathematical questions,  questioners are becoming adept at choosing those for \mo  that are amenable to its approach. The questions and the answers build upon an assumption of a high level of shared background knowledge, perhaps at the level of a PhD in group theory. A few questions were flagged in comments as inaccessible or extremely specialised parts of group theory:  we saw few questions that extended beyond the domain to link with other specialised areas of mathematics. 

 The usual presentation of mathematics in research papers is in a standardised precise and rigorous style:  for example,  the response to a conjecture is either a counterexample, or a proof of a corresponding theorem, structured by means of intermediate definitions, theorems and proofs.  By contrast, the typical response to a \mo question, whatever the category,  is a  discussion presenting facts or short chains of inference  that are relevant to the question, but may not answer it directly. The facts and inference steps  are justified by reference to the literature, or to mathematical knowledge that the responder expects the other participants to have. Thus in modelling a \mo discussion, we might think of each user as associated to a collection of facts and short inferences from them, with the outcome of the discussion being that combining the facts known to different users has allowed new  inferences. Thus the power of \mo comes from developing collective intelligence through sharing information and understanding. 

In 56\% of the responses we found citations to the literature. This includes both finding papers that questioners were unaware of, and extracting results that are not explicit in the paper, but are  straightforward (at least to experts), consequences of the material it contains. For example,  the observation needed from the paper may be a consequence of an intermediate result, or a  property of an example which was presented in the paper for other purposes.

In 34\% of the responses explicit examples of particular groups were given, as evidence for, or counterexamples to, conjectures. The role of examples in mathematical practice, for example as evidence  to refine conjectures, was explored by Lakatos \cite{lakatos1976proofs}, and \mo and {\it polymath} provide a valuable evidence base for further research \cite{Pease12}.

 In addition  \mo captures information known to individuals but not normally  recorded in the research literature: for example unpublished material, motivation, explanations as to why particular approaches  do not work or have been abandoned, and intuition  about conjectures.

The presentation is often speculative and  informal, a style which would have no place in a research paper, reinforced by conversational devices that are accepting of error and invite challenge, such as ``I may be wrong but...'', ``This isn't quite right, but roughly speaking...''.  Where errors are spotted, either by the person who made them or by others, the style is to politely accept and correct them: corrected errors of this kind were found in 37\% of our sample\footnote{This excludes ``conjecture'' questions where the responses  refutes the conjecture. We looked at {\it discussions} of error: we have no idea how many actual errors there are!} 

 It is perhaps worth commenting on things that we did not see in  our sample of technical questions tagged ``group theory'' in {\it mathoverflow}. In developing ``new'' mathematics considerable effort is put into the formation of new concepts and definitions: we saw little of this in {\it mathoverflow}, where questions are by and large focussed on extending or refining existing knowledge and theories.  A preliminary scan suggests these are not present in other technical areas of \mo either.  
 
 We see little serious disagreement in our \mo sample: perhaps partly because of the effect of the ``house rules'', but also because of the style of discussion, which is based on evidence from the shared research  background and knowledge of the participants: there is more discussion and debate in {\it meta.mathoverflow}, which  has a broader range of non-technical questions about the development of the discipline and so on. \mo seems to offer ample possibilities for further study as a exemplar of collective intelligence.
 
\section{Towards social computation}

Our long term goal is to advance mathematics through social computation, advancing current tools for example through tackling harder problems, reducing the barrier to entry of specialist knowledge,   bridging domains so mathematical specialists in different areas can collaborate, or drawing on archives of theorems and data to create new mathematics. 

%Underpinning all this will be the challenges presented in  \cite{hendler2010semantic}.

Machine resources available, in principle at least, to support  mathematics social computation, include software for symbolic and numeric  mathematics such as GAP or Maple; the mathematical literature (ambitious digitisation plans are currently being developed by the American Mathematical Society and the Sloan Foundation); and  bodies of formalised material made available from computer proofs.  The emerging field of  mathematical knowledge management \cite{kohlhase2012mathwebsearch} addresses ontologies and tools for sharing and mining such resources, for example providing ``deep'' search or executable papers. Our work on \mo highlights the importance of databases of examples, perhaps incorporating user tagging, and also of being able to mine libraries for data and deductions beyond the  immediate facts they record.

%: the work of Urban \cite{urban2006momm} on machine learning from such libraries may be significant.

Our study also highlights the importance of human factors, and of handling informal  reasoning,  error, and uncertainty.  Turning messy human knowledge into a usable information space, and reasoning across widely differing user contexts and knowledge bases is only beginning to emerge as a challenge in artificial intelligence applied to mathematics, for example in the work of Bundy  \cite{DBLP:journals/amai/Bundy11} on ``soft'' aspects such as creativity, analogy and concept formation and the handling of error by ontology repair \cite{DBLP:journals/ijswis/McNeillB07}, or work in cognitive science which studies the role of  metaphor in the evolution and understanding of mathematical concepts \cite{Lak00}. 

Social expectations in {\it mathoverflow}, and generally in research mathematics, are of a culture of open discussion,  and knowledge is freely shared provided it is attributed: for example, it is common practice in mathematics to make papers available before journal submission.  As with mathematics as a whole, information accountability in principle in a mathematics social machine comes from a shared understanding that the arguments presented, while informal, are capable of refinement to a rigorous proof. In {\it mathoverflow}, as described in \cite{tausczik2011predicting}, social expectation and information accountability are strengthened through the power of off-line reputation: users are encouraged to use real names, and are likely to interact through professional relationships beyond {\it mathoverflow}.  A further challenge for social computation will be scaling these factors up to larger more disparate communities who have less opportunity for real-world interaction;  dealing in a principled way with credit and attribution as the contributions that social computation systems make  
become routinely  significant; and incorporating models where contributions are traded rather than freely given. 

Thus mathematics offers a rich resource for the further investigation of collective intelligence, in particular the combination of precise formal deductions, and the more informal loose interaction seen in mathematical practice.

\noindent
{\bf Acknowledgements}  We acknowledge EPSRC support from EP/H500162 and EP/F02309X, and Pease in addition from EP/J004049, and   the kind hospitality of the School of Informatics at the University of Edinburgh, where this work was done while the first author was on sabbatical. We thank in particular Alan Bundy and members of the Dream group, and Robin Williams and members of the Social Informatics group, for many helpful discussions and insights.
%\end{document}  % This is where a 'short' article might terminate

%ACKNOWLEDGMENTS are optional
%\section{Acknowledgments}
%This work was partially supported under EPSRC grants ... and ..... The authors thank Andrew Aberdein, Rob Arthan, Alan Bundy, Natasa Milic-Frayling and Robin Williams for helpful discussions. 

%\begin{footnotesize}

%\bibliographystyle{abbrv}
%\begin{small}
%\bibliography{sigproc} \vspace{60 mm}
%\end{small}
%%\end{footnotesize}
%
\end{document}